# Towards a Security Lifecycle Model against Social Engineering Attacks: SLM-SEA

*Completed Research Paper*

**Tolga Mataracioglu**
TUBITAK BILGEM Cyber Security Institute
tolga.mataracioglu@tubitak.gov.tr

**Sevgi Ozkan**
Middle East Technical University
sozkan@ii.metu.edu.tr

**Ray Hackney**
Brunel University
ray.hackney@brunel.ac.uk

**ABSTRACT**

This research considers the impact of social engineering security attacks which are noted as taking opportunities for critically exploiting user awareness and behavior. The research proposes in this respect a 'managerial' method in an attempt to enhance or even ensure protection. The aim of this study is to construct a security lifecycle model against these eventualities and to analyze the test results that have been carried out within the context of the Turkish public sector. The main objective of the study is to determine why employees shared sensitive information by stating fallacies and related amendments through interviews and thus to understand user actions when they are face to face with a real social engineering attack. The research findings demonstrate that employees in Turkish public organizations are not sufficiently aware of information security and they generally ignore critically important security procedures. This represents an important illustration of the increasing need for further generalized user awareness and responsibilities where individuals and not simply software form a critical element of the security protection portfolio.

**INTRODUCTION**

Social engineering can be described as the technique of acquiring information that should not be disclosed or shared under normal circumstances through taking advantage of and using methods of influencing and convincing individuals (Mataracioglu 2010, Mataracioglu and Ozkan 2010). All employees are responsible for information security in particular the owner of the information and IT personnel. However, due to its low cost and ability to take advantage of simple technology, social engineering is frequently reported in the literature as a very effective form of attack (Winkler and Dealy 1995). About seventy percent of information theft is carried out from within the organization, either consciously or unconsciously (American Society of Industrial Security 1996, Katz 1995, Jagatic et al. 2007). In order to understand the level and extent of information security 'weaknesses' it is therefore useful to analyze user behavior.

Within the USA, according to The Federal Bureau of Investigation (FBI), companies lose nearly one hundred billion dollars per annum because of industrial espionage (Winkler 1996). The graph in Figure 1 shows a distribution, classified by breach type, of computer-related incidents that have occurred in the USA between 2002 and 2011 (DataLossDB 2012). Notably, hacking has a rate of 22%, stolen laptop with 14%, fraud with 12%, and the web with 10%. Social engineering attacks are usually performed by making phone calls or using some hardware/software tools. Evidently, if we classify the attacks made using social engineering techniques under the titles of hacking and fraud, those incidents make up a portion of crimes at a rate of 34% in the USA. According to the statistics on social engineering attacks performed in the USA during 2010, 15 companies have been contacted and a total of 135 conversations have been made (Hadnagy et al. 2010). Approximately 93% of the companies' information has been seized and most importantly only 8% of the employees actually resisted sharing their sensitive information during the phone calls.

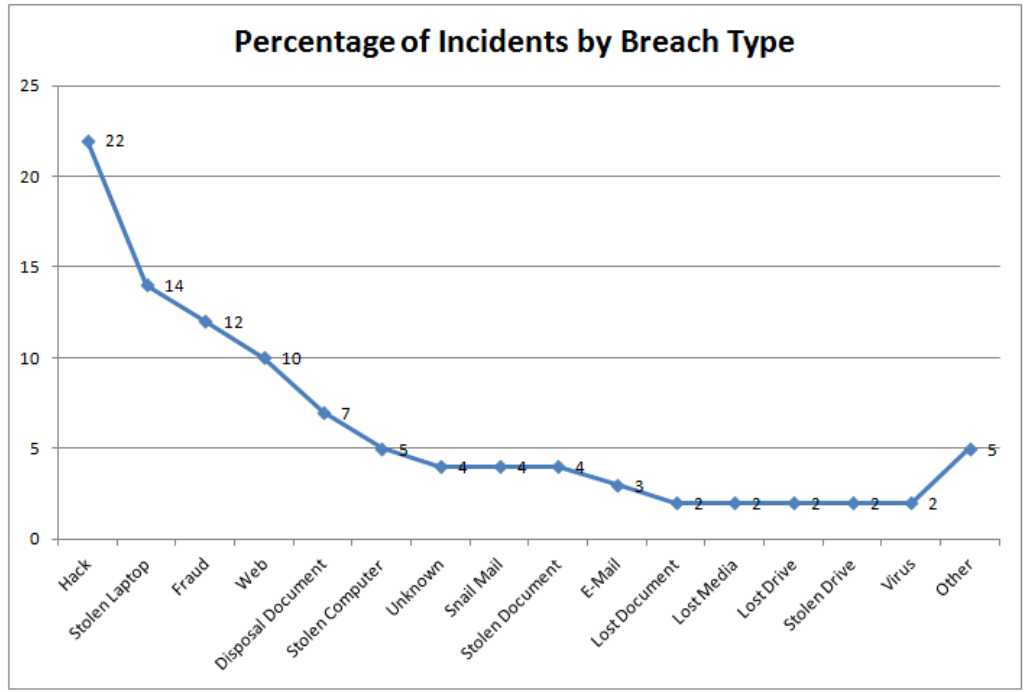

Figure 1. Percentage of incidents occured in the USA by breach type between 2002 and 2011

Most people believe they have a low chance of being deceived and being aware of this shared belief the attacker disguise their intent to arouses no suspicion and consequently exploit the victim's trust (Bican 2008, Microsoft 2010, Huber et al. 2009, Nohlberg 2008). The research in this paper provides examples of these breaches and proposes, as noted, a managerial lifecycle model in an attempt, at the very least, to reduce the impact of such attacks on sensitive information disclosure.

Social engineers considered to exist under the white hats society, welcome the information that is seemingly harmless for an organization; as it may play a crucial role in convincing others they are real (Mataracioglu 2009, Mitnick and Simon 2002, Arslantas 2004, Hasan et al. 2010). The secret of success for social engineering is that users are very much prone to being deceived if you gain their trust and if they are manipulated in a certain manner (Slatalla and Quittner 1995, Voyager 1994).

Social engineers frequently follow a certain route where the intention can be just the opposite in some cases. This is called reverse tricking (Mitnick and Simon 2002). Essentially, the attacker creates a problem where the user will be directly affected; then contact is made by telephone leaving a number for the user to call back. This so called 'penetration' is a technique where an outsider disguises as a member of organization staff to obtain passwords, etc.

Figure 2 shows the process of social engineering attacks where the first stage is research. Here, as much information about the person or organization to be attacked as possible is gathered. Second is the stage of building friendship and trust. Using the information obtained during the previous stage, the social engineer tries to gain the sympathy of the user to use this secure channel by exploiting trust to obtain sensitive information. If this information is adequate, the attack may cease. If not, the attacker goes back to the research stage and performs the cycle again.

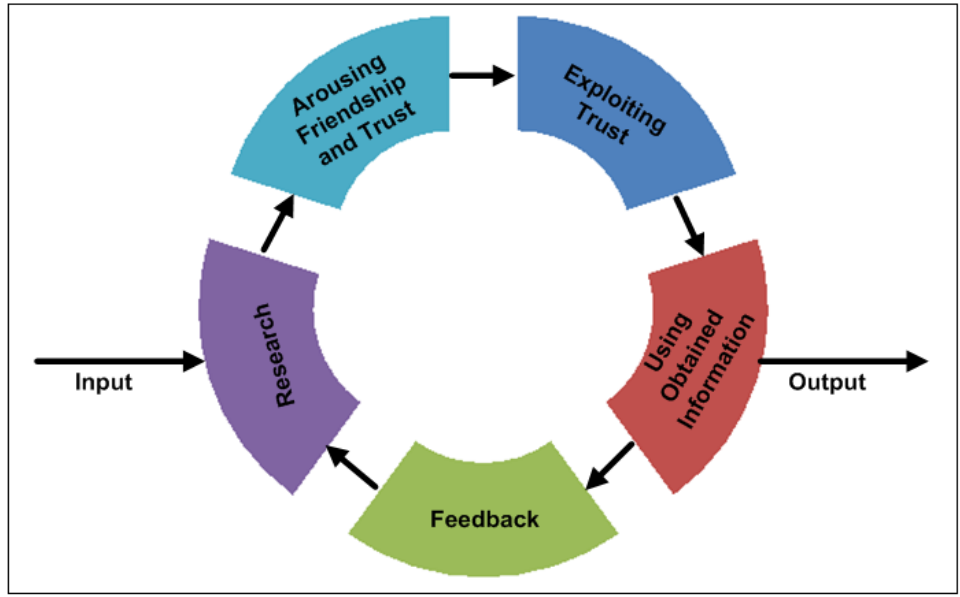

Figure 2. Social engineering process

User awareness training involving the importance of procedures and their applications, computer login and password security, hardware and software changes in the computer, laptop usage, file access and sharing, use of printers, use of portable media, virus protection, Internet security, e-mail security, back-up policy, computer security event notification, and social engineering should be given to all organization personnel in certain periods as the primary protection method (Carnegie Mellon University's Computer Emergency Response Team 2008, Rabinovitch 2007, Workman 2007, Workman 2008). As part of the "Continued Awareness Program", controls such as posting caricatures and hints, posting photos of the security personnel of the month, posting announcements on bulletin boards, posting various information security posters, sending memo emails, following security-related information on internet sites, distributing brochures and using security-related screensavers and background pictures should be included in the intranet of the organization for information security (Laribee et al. 2006, Bakhshi et al. 2009, Ceraolo 1996, Qin and Burgoon 2007).

The most frequently used methods of social engineering are to disguise as an organization employee, an employee of a company working in collaboration with the organization, an officer, a newly-recruited person or even a system producer at work to have a system patch installed (Gonzalez et al. 2006, Twitchell 2006, Major 2009). Other technical techniques include sending harmful software via email or using intra-organization terminology (Workman 2007, Workman 2008).

Our investigate implements, for clarity, some of the scenarios which have been used in social engineering attacks in public organizations for more than three years (Peltier 2006, Abawajy 2012, Jansson and Von Solms 2011, Manske 2006).

## METHOD

We formulated six different scenarios of security attacks which resulted in an approximate success rate of 70% as illustrated in Table 1. Consequently, we constructed a security lifecycle model against social engineering attacks to enable an analysis of the proposed tests that were carried out in several Turkish public sector organizations. The main contribution of the paper is the determination of and the reasons for sharing sensitive information by employees in the scenarios. The research aim was to demonstrate that information security awareness of employees is generally unacceptable. In this respect the proposed model provides valuable insights into ways of reducing compromised information security procedures.

| No | Formulated Scenarios |
|---|---|
| 1 | The data used in this scenario is obtained from outside the organization (Facebook, Google, MSN etc.). After the attacker gathers organization-related telephone information from the internet a telephone call is made suggesting that the caller is newly recruited at the IT department and a user name and password is required to update the active directory. |
| 2 | The data used in this scenario (telephone lists from the IT department, list of critical personnel etc.) is obtained from within the organization. The attacker calls these numbers again suggesting the caller is newly recruited in the IT department and is required to have a user name and password to update the active directory. |
| 3 | The attacker takes the role of an auditor and is currently in a meeting together with the CEO of the organization. A request is again made for a user name and password to enable an investigation as per the request of the CEO. |
| 4 | The data used in this scenario is obtained from outside the organization (e.g., Google). The attacker calls the telephone operator of the organization and wants to speak with someone from the accounting department. The purpose is first to obtain a name and an extension. After obtaining this information a call is made to the person in question and again attempts to elicit a user name and password. |
| 5 | If any critical software in the organization is under the operation of an external organization the attacker may suggest the role of an employee of that organization and again attempt to elicit the user name and the password for an update. |
| 6 | An email containing harmful software as attachment and having desirable content for users is prepared with the help of technical experts and is sent to users.<br>   Example 1: Please click on the attached document to see the latest salary increase table<br>   Example 2: I am a technical support group staff at the company "x". Please kindly install the attached patch on your computer for the removal of a critical gap in our company's "y" software. |

**Table 1.** Scenarios used in our social engineering attacks

Our proposed method, as a qualitative approach, against social engineering attacks is illustrated in Figure 3. We refer to the model as 'Security Lifecycle Model against Social Engineering Attacks' (SLM-SEA). The method follows the process of: first, all the organization employees receive information security awareness training after which they should be able to answer the test questions about information security awareness. The employees who have failed the test by having a score less than a certain threshold, predetermined by the organization, are sent to information security awareness training again within a short period of time. Secondly, ten sample employees, who are not previous participants, are selected for further social engineering tests.

Following the tests, the social engineer interviews the employees who have shared their sensitive information, in order to clarify user behavior. Finally, the social engineer analyzes the results and compares them with the previous results. If the percentage of the employees with acquired passwords is found to be above a predetermined threshold level, then the process will be repeated by the social engineer.

This security lifecycle fits the Plan-Do-Check-Act (PDCA) cycle which is applied in ISMS's in constructing all ISMS processes (International Organization for Standardization 2005). Further, PDCA cycle may work in every management system. In our proposed method, test phase corresponds to "plan" and "do" phases, measure phase corresponds to "check phase", and feedback and train phases correspond to "act" phase in PDCA. SLM-SEA improves users' information security awareness due to the nature of lifecycles through each iteration. Since this proposed method did not take place in technical specifications document where the phases of information security tests were written, we could not use our proposed method in the organizations that we made social engineering tests. Instead we used a subset of SLM-SEA shown in Figure 4. Compared to SLM-SEA, Plan, Do, and Check phases remain the same. However Act phase containing feedbacks and the training phases are missing in the implemented method. This works in that order: first, samples of ten employees are selected within the context of social engineering tests and the social engineer calls the samples in order to obtain their sensitive information. After those tests, social engineer interviews the employees whose sensitive information has been gathered so as to determine the user behavior. Finally, the social engineer evaluates the results.

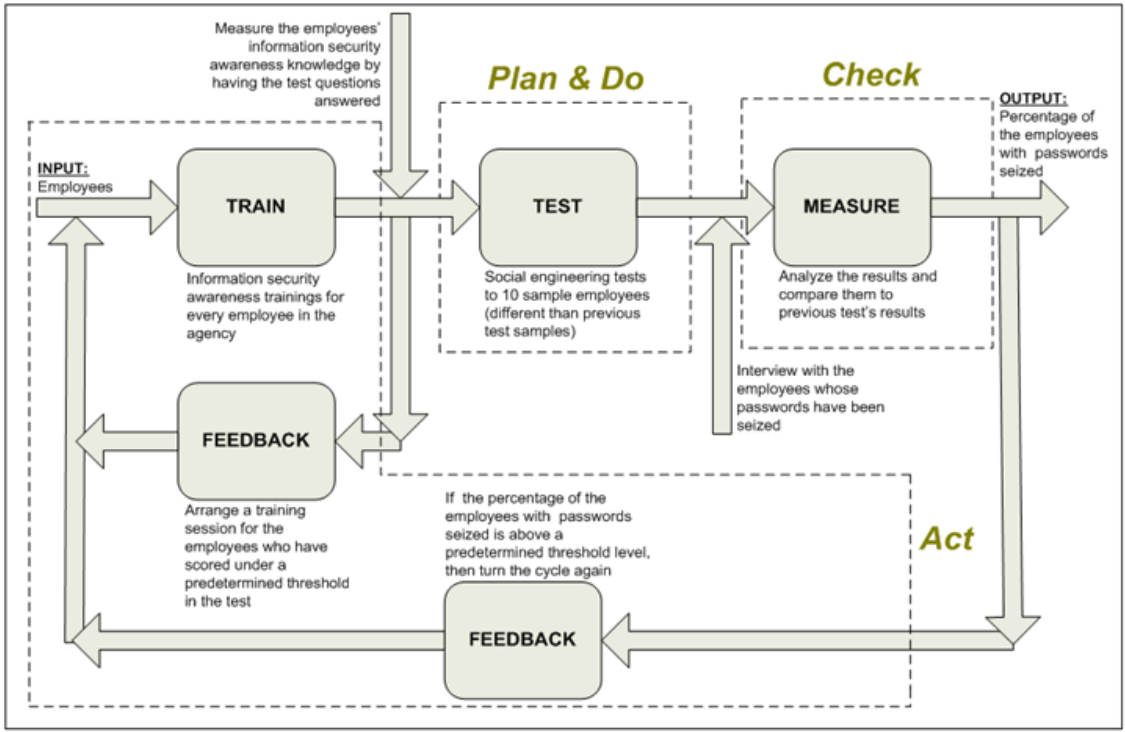

Figure 3. SLM-SEA – proposed method

Figure 5 shows the methods used at the research phase of the social engineering cycle. Web sites of the organizations were evaluated to elicit information about the organization staff, instructions and procedures. Information regarding organizations offering assistance to those organizations, if any, was also gathered. Within this process, organization-specific documents such as policies, procedures, instructions, and forms; information about organizational staff names, e-mail addresses, and telephone numbers, and information offering support to the organization are also elicited from the Internet. All this information was brought together and organization staff were called in an attempt to gather sensitive information (login passwords for computers used by organization staff or software/application passwords). The social engineer actually holds security-related certificates such as Certified Information Systems Auditor (CISA), Certified Ethical Hacker (CEH), and ISO 27001 Lead Auditor. Consequently, a member of the research team has experience and the technical ability for acquiring sensitive information from users. The social engineer is introduced, as one of the planned scenarios, as a new employee in the IT department. After the test was finished, the IT personnel were informed to ask users whose sensitive information was acquired during the test to change their passwords.

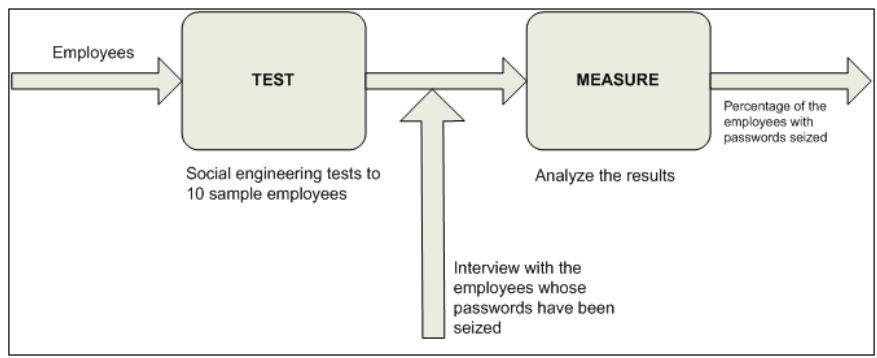

Figure 4. Implementation of SLM-SEA in our study

In the selection of sample employees in the social engineering tests, the three criteria below were considered:
1.   Sample employees were selected from critical departments (human resources, finance and accounting, sales and marketing and law.) and unit managers.
2.   After the testing of a sample was finished, another sample was selected out of the previous sample's room so as not to keep track of the tests.
3.   In every tested organization, the number of female and male sample employees was equal.

Below are the general steps followed in the simulated attacks performed in the public organizations:
1.   The social engineer gathers information about the organization, as shown in Figure 5.
2.   By using the criteria explained above, he then selects the sample employees to be called.
3.   Information from the sample employees is captured during phone calls.
4.   After the test is over, IT personnel are asked to inform the sample employees and ask them to change their passwords.
5.   The social engineer interviews the sample employees who shared their sensitive information.
6.   Finally, the social engineer analyzes the results and calculates the percentage of employees with acquired passwords and shares this information with the IT manager.
7.   Further, informs the IT manager about methods of lowering this percentage in the organization.

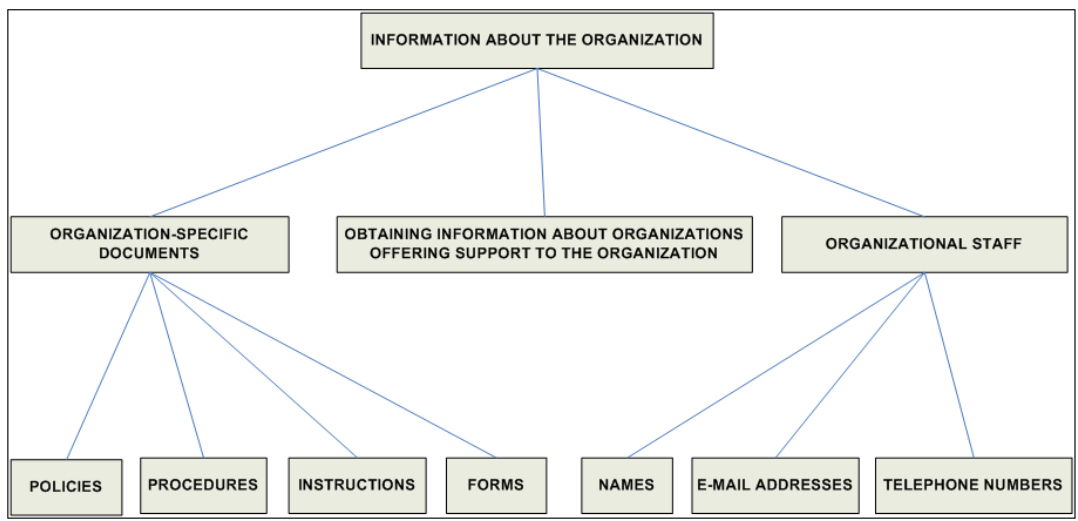

Figure 5. Information about organizations

### RESULTS

As part of the research TUBITAK BILGEM Cyber Security Institute has made social engineering attacks on six public organizations. As noted, they were Turkey's leading organizations in monetary and finance, health, telecommunications, and applied sciences. It can be reported that, as part of these tests, telephone conversations were made with 56 users and 38 passwords (corresponds to approximately 68% of users) were seized.

Table 2 contains information about tested public organizations. The organizations have been classified by their core businesses, whether ISMS has been established, employee numbers, employee types related to core businesses, and whether organizations have branch offices. As shown four different types of core business as monetary and finance, applied sciences, health, and telecommunications. Further, none of the organizations has established ISMS neither in the whole organization nor in any department of the organization. Employee numbers are approximate and differ from 100 to 800. Occupations related to core businesses of the organizations are accountants, scientists, doctors, network engineers/technicians, contractors, and recording officers. In addition, only two of the organizations related to health and monetary and finance have branch offices.

| Name | Core Business | ISMS | Employee# | Core Type of Employees | Branch Office |
|---|---|---|---|---|---|
| Organization A | Monetary and finance | No | ~200 | Accountants | Yes |
| Organization B | Applied sciences | No | ~400 | Scientists | Yes |
| Organization C | Health | No | ~150 | Doctors | No |
| Organization D | Telecommunications | No | ~800 | Network engineers/technicians | Yes |
| Organization E | Monetary and finance | No* | ~100 | Tenderees | No |
| Organization F | Monetary and finance | No* | ~300 | Recording officers | Yes |

\* ISMS has not been established when the tests were performed, but the organization obtained the ISO 27001 certificate afterwards.

**Table 2.** Tested organizations

Figure 6 illustrates the social engineering test results of organizations. The success rate (number of password-obtained participants/total number of participants) was 80% for Organization A, this rate was 50% for Organizations B and E, 60% in Organization C, 75% in Organization D and 100% in Organization F, which are seriously high. In fact, it may not be quite correct to consider success rate by organization; as it may not be possible to reach this information with this sensitive information, even if only one person's information is obtained from the organization as the attacker. When the results are analyzed, it is apparent that employees are not that aware of information security procedures (Mataracioglu and Ozkan 2010, Mataracioglu 2010). We run a one way ANOVA (ANalysis Of VAriance) so as to see if the number of passwords seized varied by organizations, and concluded that ANOVA results did not vary from one organization to another.

Table 3 notes that some sentences captured during the interviews with sample employees, especially unit managers performed after the tests and their fallacies and amendments, have been introduced and discussed. A common mistake is that most of the users believe they are actually telling the truth by saying fallacies, and corrections related to those fallacies by saying amendments. After the interviews and analyses, there exist nine fallacies in the public organizations tested in Turkey and those nine are common issues and found in all tested organizations.

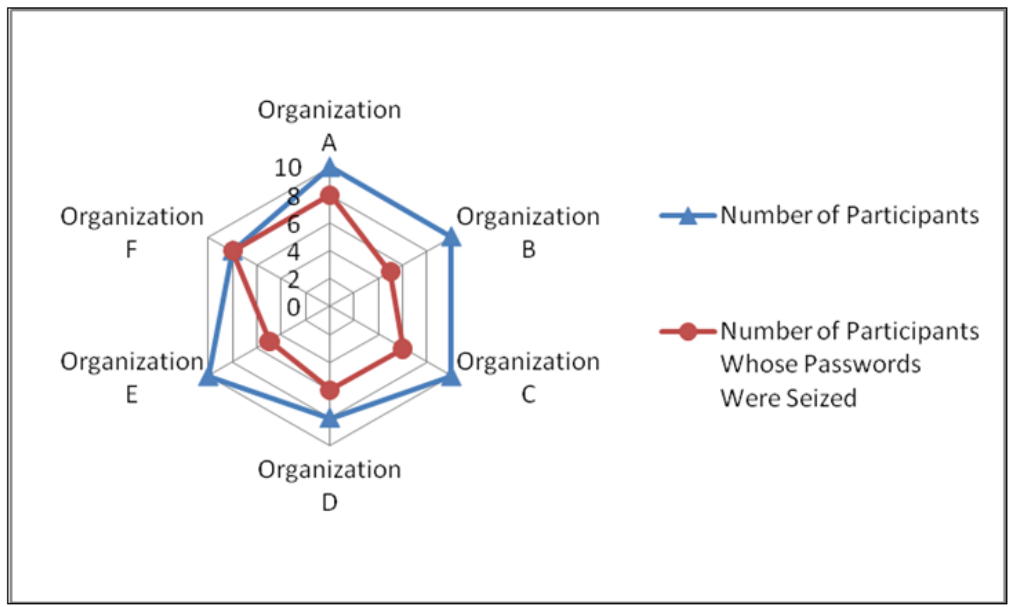

**Figure 6.** Social engineering test results

Following the data analysis we conclude our conjecture is that most of the users in organizations believe in they are aware of information security. When analyzing the social engineering results in Figure 6, we see that the performance of the organizations suffers; so we also conclude that the users in public organizations are not aware of information security. Secondly, there exists no difference between female users and male users when gathering sensitive information such as user names and passwords. As seen in Table 3, the related amendment in the first line says that only 17 female sample employees have given their sensitive information (nearly 30% of total sample employees). 21 male sample employees (nearly 70% of total sample employees) have shared their sensitive information; so we proved our hypothesis. Thirdly, all of the social engineering attacks may not come from outside of the organization; hence organizations should be precautious about the attackers from inside. We conducted our social engineering tests both in and out of the organization and concluded that the percentages of gathering sensitive information inside and outside of the organizations are approximately the same. In addition, no matter what the classification of the organization is; e.g. health, telecommunications, monetary and finance, applied sciences, user immunity against social engineering attacks does not differ if the organization has not established ISMS. Finally, comparing Figure 6 and Table 2, none of the organizations has established ISMS and the lowest success rate concerning number of participants whose passwords were seized is above 50% which is very high.

| No | Fallacy | Captured From | Related Amendment |
|---|---|---|---|
| 1 | "Female employees may share their sensitive information easier than male employees". | The organization related to monetary and finance | The results show that only 17 female sample employees have given their sensitive information (nearly 30% of total sample employees). 21 male sample employees (nearly 70% of total sample employees) have shared their sensitive information. |
| 2 | "The social engineer has called the employees by using the inhouse phones and we shared the sensitive information. If the call was external we would not have shared our sensitive information". | The organization related to monetary and finance | Sensitive information should be hidden from other employees in the organization as well as outside world. The research results show that most of the successful attacks come from inside. |
| 3 | "The social engineer is has introduced as a new employee in the organization, and we did not want to discourage him". | The organization related to health | Not giving the sensitive information to new employees is not a discouraging behavior, instead, this should be one of the clauses of the organization's information security policies and procedures. |
| 4 | During phone calls, some users asked the social engineer whether sensitive information would cause a problem. | The organization related to health | A social engineer never confesses that the release of sensitive information will cause a problem. |
| 5 | Some users were asked for IT help from the social engineer before sharing sensitive information. | The organization related to telecommunications | Sensitive information should not be shared to anyone for sake of any help or compensation. |
| 6 | "The social engineer was talking too technical during the phone call; I could not understand anything, however I shared my sensitive information". | The organization related to telecommunications | The social engineer could obtain much more technical information about the organization by using different methods stated in Part 2. |
| 7 | "The voice of the social engineer seemed to be trustworthy and I believed that sharing my sensitive information would not cause a problem". | The organization related to telecommunications | The most powerful weapon of social engineers is phone calls as they influence users most effectively |
| 8 | "I had a task to be completed as soon as possible and I shared my sensitive information with the social engineer so as to job him off". | The organization related to telecommunications | No matter what excuse is forthcoming the sensitive information should not be shared with others. |

| No | Fallacy | Captured From | Related Amendment |
|---|---|---|---|
| 9 | "The social engineer was introduced as an employee appointed under Mr./Mrs. X, the IT manager. Since the name of the IT manager is known, I trusted him working in our organization and shared my sensitive information". | The organization related to applied sciences | Knowing some employee names does not guarantee that person works for the organization. Besides, even that person works for the organization, the sensitive information should not be shared with anyone. |

**Table 3.** Fallacies and related amendments formed during post-social engineering test interviews

### DISCUSSION

Clearly, a limitation of our proposed method did not take place in technical specifications document where the phases of information security tests were written and so we could not use our proposed method in the organizations that we made social engineering tests. It is suggested that our proposed method named SLM-SEA given in Figure 3 should be entirely implemented in organizations so as to reduce the effects of social engineering attacks dramatically.

Further we only conducted our social engineering attacks in Turkish public organizations. So there is a critical need to generalize our process to other organizations including the private sector, and non-governmental organizations globally.

### CONCLUSION

The principal contribution of the paper is to determine the reasons for sharing sensitive information by stating fallacies and related amendments through interviews with employees so as to understand user behavior when a social engineering attack is carried out and to construct an information security lifecycle model against such attacks (SLM-SEA). We have found nine fallacies and stated related amendments in Table 3. The aim of the research was to establish that Turkish public organization employees lack in information security awareness and they compromise the information security principles which should be obligatory in any organization. We conclude that legislation for urging to establish ISMS, even to obtain ISO 27001 certificate in Turkish public organizations will help to turn SLM-SEA as a security lifecycle so as to immunize organizations' employees against social engineering attacks.

### ACKNOWLEDGMENT

The authors would like to thank Soner Yildirim for his valuable comments on the submitted manuscript.